\documentstyle{article}
\begin{document}
\large

\title{\bf \LARGE 
A self consistent field approach to surfaces of compressible polymer blends
}
\author
{F. Schmid \\
Johannes Gutenberg Universit\"at Mainz \\
D55099 Mainz\\
Germany
}
\date{}
\maketitle

{\bf Abstract} - 

A self consistent field theory for compressible polymer mixtures is 
developed by introducing elements of classical density functional theory
into the framework of the Helfand theory. It is then applied to study free 
surfaces of binary (A,B) polymer blends.
Density profiles in the one- and two-phase region are calculated as 
well as chain end distributions and chain orientations of the minority 
and the majority component.
In the ideally symmetric mixture, in which the individual properties of 
polymers A and B are the same and both have the same surface energy, 
polymers of the minority component segregate to the surface, where
they are exposed to less polymers of the majority component.
This effect can only be captured correctly, if one
accounts for the fact that the monomer-monomer interaction has
finite range. As a consequence, the Flory-Huggins-parameter 
varies in space and depends on the concentration profiles and their
derivatives. The surface segregation calculated with such an ansatz,
without any fit parameter, is in reasonable quantitative agreement with 
data from recent Monte Carlo simulations. 

\vfill

\newpage

\section{Introduction}

Many important properties of materials are determined by the structure
of their surfaces (adhesive properties, thin film properties etc.). 
Polymeric substances are affected by the presence of a surface 
on many different length scales: rearrangement of monomers similar to
the layering and packing of molecules at surfaces of simple liquids,
reorientation of monomers, reorientation of whole chains, 
surface induced ordering in copolymer systems, wetting and 
other surface induced transitions\cite{sanchez,binder1}.
One of the most remarkable effects at surfaces of polymeric mixtures
is the phenomenon of surface segregation - the composition of the melt 
close to the surface differs from that in the bulk. The reasons for such 
a behavior are manifold. Shorter chains tend to segregate to the surface 
because they loose the least entropy entropy
there\cite{fred1,cari,hariharan1};
Stiffer chains are favored at the surface because they pack more 
easily\cite{cari}. Hard walls may interact with the different
polymer species with different interaction strength. At free surfaces, 
segregation is induced by ``missing neighbor effects'': 
The polymer species with the lower monomer-monomer attraction segregates,
since it is less penalized for the lack of neighbors at the surface --
it's surface energy is lower\cite{hariharan2}. A similar effect occurs even in
completely symmetric mixtures of two incompatible, but otherwise 
``identical'' polymers. The minority component enriches at the 
surface, because it has less contacts with polymers of the majority
component there. This interaction driven segregation is a general 
phenomenon, which is found in many systems with a tendency of bulk
ordering (the ``ordering'' here being the demixing of the two polymer
species). It can be expected in the one-phase region as well as
in the two-phase region, and is strongest in the vicinity of the 
critical demixing point\cite{surface}. 
An additional aspect of surface segregation in polymeric systems has
to do with the competition of length scales there: 
the single chain gyration radius competes with the typical
length scale for collective concentration fluctuations, which diverges
at the critical point.

Experimental studies have mainly been concerned with surface segregation
in polymer mixtures in the one-phase region\cite{sokolov}-\cite{tasaki}. 
Popular systems are, for example, mixtures of polystyrene (PS) and
deuterated polystyrene (dPS), or of hydrogenated and deuterated
polymethylmethacrylate (PMMA and dPMMA).
The surface of PMMA/dPMMA mixtures saturates with dPMMA almost independently 
of chain lengths and bulk composition\cite{tasaki}. 
Hence the difference between the surface energies of the components is 
presumably very high and entirely dominates the segregation process. 
The segregation tendency of dPS in PS/dPS mixtures is much weaker and the
dPS excess at the surface depends strongly on the bulk volume fraction. The 
relative importance of different segregation mechanisms
can be estimated by simple considerations. 
The excess surface energy of dPS at vanishing bulk volume fraction 
is roughly $\mu \approx 2-3 \cdot 10^{-3} nm$\cite{jones1} at temperature 
$184^0C$.  The Flory Huggins parameter $\chi$, which measures the relative 
repulsion of monomers of the two components, takes the value 
$\chi = 1.5\cdot10^{-4}$, and the polymer segment length is $a=0.67 nm$.
From comparing $\mu$ to $\chi a$, one concludes that
the segregation process is still mainly driven by the excess 
surface energy, but that interaction driven segregation may become important. 
An even stronger effect can be expected in mixtures where $\chi$ is larger.

Apart from experiments, Computer simulations provide additional understanding
and a good testing ground for theories\cite{binder2}. Monte Carlo simulations 
of simple one component polymer melts at surfaces have given 
insight into the structure of polymer chains close to the surface -- chain 
end distributions, chain orientations, etc.\cite{dickman2}-\cite{wang1}.
Being  computationally very demanding, simulations of polymer 
blends at surfaces are still scarce. 
Wang et al have investigated the wetting behavior of very short 
chains\cite{wang2} at walls with very strong attraction of one species. 
In a recent study, Rouault et al\cite{rouault} consider incompatible symmetric 
polymer blends at ``neutral'' walls and find, as expected, segregation of 
the minority component there. In contrast, Cifra et al\cite{cifra} study 
fully compatible blends, where the missing neighbor effect leads to the 
enrichment of the majority component at the surface.

Theoretical treatments of polymeric surfaces have often dealt with 
simple Flory-Huggins-de Gennes functionals of the concentration profiles
in ``square gradient'' approximation\cite{nakanishi}-\cite{flebbe}.
Jones et al\cite{jones2,jones3} first pointed out that this type of 
theory fails to reproduce the correct form of segregation profiles: Unlike 
profiles typically obtained from Ginzburg-Landau theories, experimental 
profiles are flattened right at the surface. More sophisticated mean field 
approaches take the chain connectivity into account more explicitly,
either in the framework of integral equation theories or density functional 
theories\cite{woodward}-\cite{coy2}, or in the self consistent field theory 
originally developed by Helfand and others\cite{helfand1}-\cite{shull},
or, lately, in combinations of the two approaches\cite{freed3,coy2}.
A recent study by Genzer et al\cite{genzer} has shown that profiles of polymer
blend surfaces calculated within the self consistent field theory can 
indeed be fitted to experimentally measured profiles in a satisfactory way. 
However, Genzer et al treat the blend as an incompressible mixture, a
simplification which is clearly not justified when treating
free surfaces. Surface segregation is enforced by external ad hoc potentials 
acting on the surface layer. The theory does not allow for a microscopic 
treatment of missing neighbor effects.

The importance of compressibility effects in polymer melts has been
recognized by several authors. In the early work of Helfand et al,
they are taken into account within a quadratic approximation. This is
reasonable as long as one deals with small total density fluctuations, not 
in the vicinity of a surface where the density goes down to zero. 
Some work has been devoted to elaborated density functional theories of one 
component polymer melts at surfaces\cite{woodward,kierlik,mcmullen,coy1}. 
Those studies have had remarkable success in reproducing the density profiles 
measured in corresponding Monte Carlo Simulations; however, they mainly 
focussed on the microscopic structure of a melt over the length scale of 
the monomer size -- here, we are interested in composition variations
on a larger scale. Freed and coworkers have developed density functional 
theories of compressible polymer blends\cite{freed2}. Recently, Hariharan and 
Harris have presented a self consistent field study of the surface of a 
compressible copolymer melt within the framework of the Scheutjens and 
Fleer lattice theory\cite{scheutjens,hariharan3}. 
Compressibility can be introduced into 
lattice models in a relatively straightforward way by admitting vacancies
as a new type of ``particles'' which may occupy a lattice site. To
the present author's knowledge, no attempt has been made so far to study
surfaces of compressible polymer mixtures in continuum space.

This is the scope of the present work. On the base of the equation of
state which is assumed to be known and using elements of simple
density functional theory, a self consistent field theory
for compressible polymer melts will be developed. In self consistent
field theory, monomer-monomer interactions are commonly treated as
contact interactions $V(\vec{r}) \propto \chi \delta (\vec{r})$. 
Such a potential Ansatz is not suitable for the description
of important missing neighbor effects -- e.g. the effect that, close
to a hard wall, monomers a $z>0$ have no neighbors at $z<0$.
Hence the theory will be extended such that it accounts for the finite range
of interactions. 
The parameters of the theory will be adapted to the bond-fluctuation model 
used in the Monte Carlo simulations of Rouault et al\cite{rouault}. 

The bulk properties of the model are very well known from previous
work\cite{bfm,marcus1}. M\"uller et al performed large
scale simulation studies of interfacial properties in incompatible polymer 
mixtures\cite{marcus2}. 
The comparison of his results to self consistent field 
calculations showed that the theory is rather successful in predicting 
many quantities even at the relatively short chain length of $N=32$, 
except for the interfacial width\cite{schmid}. The discrepancies between
theory and simulation becomes smaller as the chain length is increased,
but the convergence is extremely slow.
Similarly, we will see here that our self consistent field calculation
yield the quantitatively correct values of the surface segregation,
without any fit parameter, but fail to predict the correct decay
lengths of the profiles. 

The paper is organized as follows. The theory is developed and the model 
parameters are determined in the next section.
In section 3, the surface segregation at neutral walls
is calculated in the one-phase and in the two-phase region and
compared to the available simulation data. The influence
of chain length asymmetry on the surface segregation is also studied.
Then the structure of the blend at the surface is 
analyzed in more detail. The evolution of density 
profiles as one moves away from the critical point will be examined,
chain end distributions and chain orientations will be calculated.
The results are summarized in section 4.

\bigskip

\section{Self consistent field theory for compressible melts}

\bigskip

We consider mixtures of flexible polymers $A$ and $B$ in the vicinity of a 
surface. Each chain of type $i$ ($i$ = $A$ or $B$) contains $N_i$ monomers 
and is characterized by a statistical segment length $b_i$, The radius
of gyration of a chain in the homogeneous melt is thus given by
$R_{g,i}{}^2 = b_i{}^2 N_i/6$ (random walk).
Unless stated otherwise, we will assume that polymers $A$ and $B$ have
the same properties, i.e. $N$, $b$ and the bulk density in
a pure system $\rho_b$ are equal for both species.

Molecules are treated as space curves $\vec{r}(s)$, with $s$ varying 
from 0 to 1. The partition function for a system of $n_A$ polymers of type 
$A$ and $n_B$ polymers of type $B$ has the general form\cite{helfand2}
\begin{equation}
\label{part}
{\cal Z} = \frac{1}{n_A! n_B!} \: \int
\prod_{i=1}^{n_A} \widehat{\cal D} \{ \vec{r}_i(\cdot) \}
\prod_{j=1}^{n_B} \widehat{\cal D} \{ \vec{r}_j(\cdot) \}
\exp[ - \beta {\cal F} \{ \widehat{\rho}_A, \widehat{\rho}_B \}],
\end{equation}
where $\beta = 1/k_B T$ is the Boltzmann factor and
$\widehat{\rho}_A$, $\widehat{\rho}_B$ are monomer density
operators
\begin{equation}
\widehat{\rho}_{A,B}(\vec{r}) = N \:
\sum_{i=1}^{n_{A,B}} \int_0^1 ds \: \delta(\vec{r}-\vec{r}_i(s))
\end{equation}
and ${\cal F}\{\widehat{\rho}_A,\widehat{\rho}_B\}$ represents a
coarse grained free energy functional, which will be discussed in more 
detail below. Individual space curves in the functional integral are 
assigned a statistical weight $\widehat{\cal D} \{ \vec{r}(\cdot) \} =
{\cal D} \{ \vec{r}(\cdot) \} \: {\cal P}_G \{ \vec{r}(\cdot) \} $
with
\begin{equation}
{\cal P}_G \{ \vec{r}(\cdot) \} = {\cal N}
\exp[ - \frac{3}{2 N b^2} \int_0^1 ds | \frac{d\vec{r}}{ds}|^2 ],
\end{equation}
with the normalization factor ${\cal N}$. Flexible chains are hence modelled 
as continuous space curves, which are distributed according to a Gaussian 
stretching energy with spring constant $3/N b^2$. 

We treat the problem in mean field approximation by replacing the monomer
density operator $\widehat{\rho}(\vec{r})$ in eqn (\ref{part}) with the
average monomer density 
\mbox{$\rho_i(\vec{r}) := \langle \widehat{\rho}_i(\vec{r}) \rangle$}.
Monomer-monomer correlations are ignored. The chains can then be
treated as independent random walks in the external field
$W_i(\vec{r}) = \frac{\delta (\beta {\cal F})}{\delta (\rho_i(\vec{r}))}$
\cite{helfand2}. It is useful to define the end-segment distribution functions
\begin{equation}
Q_i(\vec{r}_0,s) = \int \widehat{\cal D} \{ \vec{r}(\cdot) \}
\exp [ - \int_0^s ds' W_i(\vec{r}(s'))]  
\delta(\vec{r}_0 - \vec{r(s)}) 
\end{equation}
which obey the diffusion equation\cite{helfand1}
\begin{equation}
\label{qq}
\{ \frac{1}{N_i} \: \frac{\partial}{\partial s} 
- \frac{1}{6} b^2 \vec{\nabla}_{\vec{r}}^2 + W_i \} \; Q_i(\vec{r},s) = 0 
\end{equation}
with initial condition $Q_i(\vec{r},0) \equiv 1$. The function $Q_i$ gives 
the statistical weight for a part of a polymer of type $i$ and length $sN$, 
with one end fixed at position $\vec{r}$. The average
density of type $i$ monomers (with orientation $\vec{u}$) is then given by 
\begin{equation}
\label{phi}
\rho_i(\vec{r}) = \rho_b \int_0^1 ds \; Q_i(\vec{r},s) Q_i(\vec{r},1-s),
\end{equation}
which completes the cycle of self-consistent equations. The calculated
profiles can be used to calculate the configurational entropy of a polymer
of type $i$
\begin{equation}
{\cal S}_i = k_B (\log {\cal Z}_i +
\frac{1}{n_i} \int d \vec{r} W_i(\vec{r}) \rho_i(\vec{r}) ),
\end{equation}
where the single chain partition function 
${\cal Z}_i = \int d\vec{r} Q_i(\vec{r},1) = N n_i/\rho_b$ is a constant.
From there one gets the free energy
\begin{equation}
\beta F = \beta {\cal F}\{\rho_A,\rho_B\} - \frac{1}{k_B}
 (n_A {\cal S}_A + n_B {\cal S}_B).
\end{equation}

We now turn to the discussion of the free energy functional
${\cal F}\{\widehat{\rho}_A,\widehat{\rho}_B\}$, which defines the
concrete model. We assume that species dependent interactions between
monomers can be treated as perturbations of a reference system
``0'', in which there is no distinction between monomers of different
type. 
\begin{equation}
{\cal F}\{\rho_A,\rho_B\} = 
{\cal F}_0\{\rho\} + {\cal F}_{inter}\{\rho_A,\rho_B\} 
\qquad \mbox{with} \qquad
\rho = \rho_A + \rho_B
\end{equation}
More specifically, our Ansatz is based on a reference system of 
noninteracting polymers with simply hard core repulsion. We
can treat the reference system in local density approximation
\begin{equation}
\beta {\cal F}_0\{\rho\} 
= \int d \vec{r} \rho(\vec{r}) (f[\rho(\vec{r})] - \mu).
\end{equation}
The bulk free energy per monomer $f(\rho)$ can be calculated
from the equation of state $\Pi(\rho)$ (with the pressure $\Pi$) using
\begin{equation}
f(\rho) = \int_0^{\rho} dx \: \Pi(x)/x^2;
\end{equation}
the chemical potential $\mu$ drives the total number of particles and is
related to the bulk density $\rho_b$ via
$\mu = f(\rho_b) + \rho_b \frac{\partial f}{\partial \rho_b}$.
We will use the equation of state resulting from the 
generalized Flory theory in continuous space proposed by Dickman and 
coworkers\cite{dickman,hpd}. 
\begin{equation}
\label{eos}
\Pi = C_0 \rho [\frac{1+\eta+\eta^2-\eta^3}{(1-\eta)^3}-1],
\end{equation}
for long enough chain lengths.
The constant $C_0$ is $C_0=0.6583$, and the ``packing fraction'' $\eta$
is given by $\eta = a^3 \rho \pi/6 $ with the ``monomer diameter'' 
$a=1.96 a_0$ in the bond fluctuation model\cite{hpd} (where $a_0$ is the 
lattice spacing).
This theory has been remarkably successful in predicting the correct 
pressures for the athermal bond-fluctuation model over a wide range of 
densities. At the interesting density $\rho=1/16 a_0^{-3}$, however,
it yields a compressibility $\kappa =34 a_0^3$ which is slightly too high
($\kappa \approx 31 a_0^3$ according to simulations\cite{marcus1}). In order
to correct for this, a higher order term is added to $f(\rho)$,
\begin{equation}
\label{eos1}
f(\rho) = C_0\frac{\eta (4-3 \eta+ b \eta^3)}{(1-\eta)^2} 
\qquad \mbox{with} \qquad b=3.0 .
\end{equation}
The ad-hoc correction to the generalized Flory theory ($b=0$) 
is motivated by the fact that the equation of state derived from
(\ref{eos1}), $\Pi = \rho^2 f'(\rho)$, is numerically almost
identical with (\ref{eos}) in the relevant density region
$0 \le \rho \le a_0^3/16$. Calculations have also been performed
at $b=0$ or with a completely different form of $f(\rho)$ (a quadratic
expansion around $\rho=a_0^3/16$, cf. \cite{schmid}) -- the results
appear to be rather insensitive to details of the equation of state.

The formalism so far takes care only of the hard core repulsion between 
monomers. In a perturbative treatment, monomers $A$ and $B$ are assumed to 
interact with each other with additional integrable potentials 
$U_{AA}(\vec{r})$, $U_{BB}(\vec{R})$ and $U_{AB}(\vec{r})$, and the energetic 
contribution to the free energy is given by
\begin{equation}
\label{fi1}
{\cal F}_{inter} = \int d \vec{r} d \vec{r}'
\frac{1}{2} \bigg[ \sum_{i,j} \rho_{ij}^{(2)}(\vec{r},\vec{r}') 
U_{ij}(\vec{r}-\vec{r}') \bigg],
\end{equation}
where the indices $i,j$ run over the polymer species $A$ and $B$, and
$\rho_{ij}^{(2)}(\vec{r},\vec{r}')$ is the pair density of type $i$ monomers 
at point $\vec{r}$ and type $j$ monomers at point $\vec{r}'$. We use the 
simple approximation 
\begin{equation}
\label{pair}
 \rho_{ij}^{(2)} (\vec{r},\vec{r}')
= \gamma(\vec{r}-\vec{r}') \rho_i(\vec{r}) \rho_j(\vec{r}'),
\end{equation}
where $\gamma(\vec{r})$ does not depend on densities any more, but does
account for local effects of chain connectivity. A possible choice is
$\gamma(\vec{r}) = 0$ inside the excluded volume of the central monomer, and
$\gamma(\vec{r}) = 1 - p(\vec{r})$ outside, where $p(\vec{r})$ is the
probability that direct neighbor monomers along the chain block
point $\vec{r}$ for occupation. The Ansatz (\ref{pair}) is chosen in the
usual spirit of Flory Huggins type theories, where monomers are subject to
averaged fields created by {\em all} other monomers, and the effect of long 
range concentration correlations induced by chain connectivity is ignored.
This implies that the demixing behavior is governed by the {\em inter}chain 
interactions, as has indeed been found in simulations, see ref. \cite{marcus3}. 
Hence the pair correlation function $\gamma(\vec{r})$ is taken to be 
independent of the identity of the monomers. Note that, as soon as intrachain
interactions become important, demixing goes along with the collapse of
polymers and the Helfand treatment of chains as random walks is no longer
appropriate.

Equation (\ref{fi1}) can now be rewritten as
\begin{displaymath}
\beta {\cal F}_{inter} = \int d \vec{r} d \vec{r}' \gamma(\vec{r}-\vec{r}')
[\rho(\vec{r}) \rho(\vec{r}') V_1(\vec{r}-\vec{r}')
+ \rho_A(\vec{r}) \rho_B(\vec{r}') V_2(\vec{r}-\vec{r}')
\end{displaymath}
\begin{equation}
\label{fi2}
+ (\rho_A(\vec{r})-\rho_B(\vec{r})) \rho(\vec{r}') V_3(\vec{r}-\vec{r}') ]
\end{equation}
with
\begin{eqnarray*}
V_1(\vec{r}) &=& \frac{\beta}{4} (U_{AA}(\vec{r}) + U_{BB}(\vec{r})) \\
V_2(\vec{r}) &=& \frac{\beta}{2} 
(2 U_{AB}(\vec{r}) - U_{AA}(\vec{r}) - U_{BB}(\vec{r})) \\
V_3(\vec{r}) &=& \frac{\beta}{4} (U_{AA}(\vec{r}) - U_{BB}(\vec{r})).
\end{eqnarray*}
The first term depends only on the total density profile $\rho(\vec{r})$ 
and does not introduce qualitative changes compared to the reference
state ${\cal F}_0$. The second term describes effective
interactions between monomers $A$ and $B$, and the third term an
effective field favoring one of the components. One can see this from
looking at the homogeneous bulk system. For $i=1,2,3$ we define
\begin{equation}
\label{ui}
u_{b,i}=\rho_b \int d \vec{r} \gamma(\vec{r}) V_i(\vec{r}). 
\end{equation}
The contribution (\ref{fi2}) to the bulk free energy is then given by 
\begin{equation}
\beta {\cal F}_{inter,b} = u_{b,1} n + u_{b,2} n_A n_B/n + u_{b,3} (n_A-n_B),
\end{equation}
with the total number of monomers $n_A$, $n_B$ and $n=n_A+n_B$. Hence
we can identify $u_{b,2}$ with the Flory Huggins parameter $\chi_b$,
and $u_{b,3}$ with an effective chemical potential difference $h_b$.

We assume that the potentials $U_{ij}$ are short range and expand the 
profiles $\rho_{A,B}(\vec{r}')$ around $\vec{r}$.
Since we study a planar surface, the system is inhomogeneous only in one
space variable $z$. 
Using $\int d\vec{r} \gamma(\vec{r}) V_i(\vec{r}) z = 0$ (due to the
inversion symmetry of $V$ and $\gamma$), and defining
\begin{equation}
\label{ki}
k_i=\frac{\rho_b}{2 u_{b,i}}\int d\vec{r}\gamma(\vec{r})V_i(\vec{r}) z^2 ,
\end{equation}
one can reexpress terms of the form
\mbox{$\int d\vec{r} d\vec{r}' \rho_{\alpha}(\vec{r}) \rho_{\beta}(\vec{r}')
V_i (\vec{r}-\vec{r}') \gamma(\vec{r}-\vec{r}')$} by
\mbox{$\int dz \rho_{\alpha}(z) (\rho_{\beta}(z) + \rho_{\beta}''(z) k_i)
u_{b,i}/\rho_b$}, where $\alpha,\beta= A$ or $B$.
One obtains the total free energy\cite{comment}
\begin{eqnarray}
\label{free}
\beta {\cal F} & = &
\int dz \{ \rho(z) (f[\rho(z)]-\mu) + 
\frac{u_{b,1}}{\rho_b} \rho(z) (\rho(z) + k_1 \rho''(z)) \nonumber\\
&& + \chi(z) \rho_A(z) \rho_B(z)/\rho_b + h(z) (\rho_A(z)-\rho_B(z)) \}
\label{fi3}
\end{eqnarray}
with the space dependent Flory-Huggins parameter
$\chi(z) = \chi_b (1+\frac{1}{2} k_2 
(\frac{\rho_A''}{\rho_A} + \frac{\rho_B''}{\rho_B}))$ and the space
dependent field $h(z) = h_b (\rho + k_3 \rho'')/\rho_b$. The
difference to the bulk values of $\chi$ and $h$ reflects the
missing neighbor effects and is ultimately responsible for the
occurrence of segregation. This is obvious for the field $h(z)$ -- 
as we will see, it is also true for $\chi(z)$.
From eqn (\ref{fi3}) one can calculate the mean fields
\begin{eqnarray}
\label{fields}
W_A(z) &=& \zeta(z) + \chi_b (\rho_B + k_2 \rho_B'')/\rho_b + h(z) \\
W_B(z) &=& \zeta(z) + \chi_b (\rho_A + k_2 \rho_A'')/\rho_b - h(z) \nonumber
\end{eqnarray}
with
\begin{equation}
\zeta(z) \approx \rho \frac{df}{d\rho} + f(\rho) - \mu + \zeta_{inter}(z),
\end{equation}
\begin{displaymath}
\zeta_{inter}(z) = 
2 (\rho + k_1 \rho'') \frac{u_{b,1}}{\rho_b} + 
((\rho_A-\rho_B) + k_3 (\rho_A''-\rho_B'')) \frac{h_b}{\rho_b}
\end{displaymath}
We will assume that the perturbation ${\cal F}_{inter}$ is small
compared to ${\cal F}_0$ and that the corresponding contributions to
$\zeta(z)$, $\zeta_{inter}(z)$, can be neglected.

To summarize, elements of a simple classical density functional theory 
(the local density approximation for the treatment of hard core monomer
interactions in connection with perturbative treatment of attractive
interactions) have been introduced into the Helfand formalism, in order
to enable the study of compressible systems with intermonomeric interactions
of finite range. A similar approach has been taken by S.K. Nath et al in order 
to study surfaces of homopolymer melts\cite{coy2}. We note that it has some
inconsistencies, polymers are treated as gaussian strings, whereas monomers 
are assumed to have finite size. This should however be of little importance, 
as long as we are interested in density variations on scales larger 
than the monomer size, and as long as the chains are much longer than the 
monomers.

The model parameters are adjusted to fit the bond fluctuation model,
with the model parameters used by Rouault et al in their Monte
Carlo simulation\cite{rouault}. The bond fluctuation model is a lattice
model, where monomers occupy each a cube of eight neighboring sites
and are connected by bond vectors of length $2 \le d \le \sqrt{10}$
lattice spacings. Thus the excluded volume of a monomer is $V_{excl}=27 a_0^3$.
Simulations are often carried out at monomer density $\rho = 1/16 a_0^{-3}$, 
where one half of all available lattice sites are occupied and
the model reproduces many properties of dense polymer melts. 
Rouault et al simulated polymers of length 32 and used the interaction 
potentials $U_{AA}(\vec{r})=U_{BB}(\vec{r})=-U_{AB}(\vec{r})=-k_BT \epsilon$
for $|\vec{r}| \le \sqrt{6}$, i.e.  they have $V_1=- \epsilon/2$, 
$V_2 = 2 \epsilon$ and $V_3 = 0$ within the interaction range. No effective
field $h(z)$ is induced, since the interaction parameters are symmetric
with respect to $A$ and $B$. The sites in the interaction region
$2 \le | \vec{r} | \le \sqrt{6}$ of a monomer are at the relative
positions $\vec{r} = (200), (210), (211)$, plus permutations
of the coordinate axes and sign reversals. Hence the interaction region
covers $N_t=54$ sites, $N_2=18$ of which are located at distances $z=\pm 2$
from the central site along the $z$-direction, and $N_1=24$ at distances 
$z=\pm 1$. Assuming that a site in the interaction range of a monomer
is blocked by one of its neighbor monomers along the chain with
the probability $p=1/3$, a rough estimate of the function $\gamma(\vec{r})$ 
yields $\gamma(\vec{r})=1-p=2/3$ in the interaction region and therefore
$\chi_b = V_2 z_{eff}$ with the effective coordination number
$z_{eff} = (1-p) N_t \rho_b $ with $\rho_b = 1/16$ (cf. eqn.(\ref{ui})).
The value $z_{eff}=2.25$ calculated from this approximation is reasonably 
close to the average number of interchain contacts found in simulations, 
$z_{eff}=2.65$ at chain length 32\cite{marcus2}.  
Similarly, one estimates from eqn (\ref{ui}) and eqn(\ref{ki})
$k_2= \sum_j j^2 N_j /(2 N_t) = 8/9$.

It should be emphasized that the theory was adjusted to the 
bond fluctuation model, a lattice model, for the sole reason that simulation 
data are available, which permit to test the theoretical predictions. 
Adjusting the theory to continuous off-lattice models causes no conceptual
difficulties in most cases. It requires the knowledge of the equation of 
state in a reference system of identical polymers, a reasonable guess of the 
pair correlation function $\gamma(\vec{r})$, and the knowledge or a good 
ansatz for additional integrable intermonomeric interactions $U_{AA}, U_{AB}$ 
and $U_{BB}$, which are small enough to be treated perturbatively. Extensions
of the theory will be necessary if the origin of the Flory-Huggins parameter
$\chi$ is mostly entropic, e.g. if polymers $A$ and $B$ demix as a result of 
different monomer sizes.

A surface at $z=0$ is imposed by requiring $\rho_i(z) \equiv 0$ for $z < 0$; 
at the bulk side, the boundary conditions 
$\frac{\partial}{\partial y} Q_i(z_{max},s) = 0$ and
$W_i(z_{max}) = W_{i,bulk}$ were used, where $W_{i,bulk}$ denotes the 
bulk value of the fields $W_i$. The numerical method used to find the
self consistent solution of the diffusion equation was an iterative
relaxation technique described in ref. \cite{schmid}. 
The iteration cycle was stopped after reaching an accuracy 
$\Gamma = {\sum_i \int dy (\Delta W_i)^2}$ of $10^{-14}$.
(relative accuracy $\ge 10^{-15}$).

\bigskip

\section{Results}

\bigskip

The system shows the usual bulk demixing phase behavior.
From eqns (\ref{qq}) and (\ref{phi}), one can see that the bulk volume 
fractions $\Phi_{A,B}=\rho_{A,B}/\rho_b$ obey the mean field equations 
$\Phi_i = \exp(-N_i W_i)$ or, equivalently,
$\partial F/\partial \rho_i = \log (\Phi_i)/N_i + W_i = 0$
with the Flory type bulk free energy per volume
\begin{equation}
\beta F = (\rho_A + \rho_B) f(\rho_A + \rho_B) + 
\sum_i \frac{\rho_i}{N_i} \log(\Phi_i) 
+ \frac{\chi_b}{\rho_b} \rho_A \rho_B - \sum_i H_i \rho_i,
\end{equation}
and $H_A = \mu + h_b + 1/N_A, H_B = \mu - h_b + 1/N_B$. 
It is useful to rewrite the free energy as a function of the total density 
$\rho = \rho_A + \rho_B$ and the density difference $m=\rho_A - \rho_B$. 
A homogeneous blend is stable or metastable with respect to small
fluctuations in $m$ as long as 
$[\partial^2 F/(\partial m)^2]_{\rho} = (\chi_s -\chi)/(2 \rho_b) > 0$
with the spinodal $\chi_s = (1/(N_A \Phi_A) + 1/(N_B \Phi_B))/2$.
The spinodal ends in the critical point, which is characterized by 
$[\partial^2 F/(\partial m)^2]_{\rho} = 0$
and $[\partial^3 F/(\partial m)^3]_{\rho} =
((1/(N_A \Phi_A^2) - 1/(N_B \Phi_B^2))/(2 \rho_b) =0$.
Hence one gets the critical $\chi$ parameter and critical volume fractions
\begin{equation}
\label{crit}
\Phi_{A,c} = \frac{\sqrt{N_B}}{\sqrt{N_A}+\sqrt{N_B}} \qquad
\Phi_{B,c} = \frac{\sqrt{N_A}}{\sqrt{N_A}+\sqrt{N_B}} \qquad
\chi_c = \frac{1}{2}(\frac{1}{\sqrt{N_A}} + \frac{1}{\sqrt{N_B}})^2
\end{equation}
At low temperatures or $\chi > \chi_c$, the mean field equations have
no stable homogeneous solution for a range of concentration differences $m$,
i.e. the system demixes. This implies that 
there exist two distinct solutions in a range of $h_b$, 
one finds metastable states or two phase coexistence;
at $\chi_c$ the two solutions merge continuously into one.
The order parameter of the system can be defined as the difference of
volume fractions in the two coexisting phases.
In the symmetric system with equal chain lengths $N_A=N_B$, two-phase
coexistence occurs below the critical point $(\chi N)_c = 2$ at
$h_b=0$ (see Fig. 1).

\subsection{Surface segregation}

Figure 2 shows a segregation profile in a symmetric mixture at bulk 
two phase coexistence. The volume fraction of the minority component is 
increased almost by a factor of two at the surface. After a
flatter part, the composition profile decays towards 
the bulk value within $2-3$ times the gyration radius $R_g$. The total
density profile is shown in the inset: The density reaches the bulk value much
more quickly than the composition profile. This reflects the 
different length scales involved in the problem: The relaxation of the
total density profile involves rearrangement of single "monomers", it is
hence driven by the Kuhn length $b$. The composition change at the surface, 
on the other hand, goes along with the segregation of whole chains -- the
thickness of the segregation layer has to be of the order of at least the
gyration radius $R_g$. The fine structure of the profiles will be discussed
in more detail later. It is worth noting that the extent of surface 
segregation strongly depends on the range of the monomer interactions or 
the value of $k_2$. In the limit of $k_2=0$ or pure contact interactions, 
segregation is almost entirely suppressed (dashed line).

Apart from the bulk phase diagram, Figure 1 also indicates the surface
volume fraction of component $A$ in the different phases as a function
of ``temperature'' $1/\chi N$ at two-phase coexistence. The difference
$\Delta \Phi_A$ between surface and bulk volume fraction predicted by the 
theory is compared to the simulation results of Rouault et al in Fig. 3. 
(Note that our mean field theory overestimates the critical temperature and 
does not describe correctly the Ising-type critical behavior found in 
short chain systems\cite{bfm,marcus2}, but the region near criticality is not
our main interest here.)
The simulations were performed in a slab geometry with a slab thickness of
around three times the gyration radius. In such a thin film, the 
critical temperature is reduced, and so is the order parameter in the two
phase region. The difference between the surface volume fraction of the 
minority phase in the thin film and the bulk volume fraction in an 
``infinite'' system can hence be taken as an upper bound for
$\Delta \Phi_A$. A lower bound is given by the difference between the
volume fraction at the surface and the center of the film. Figure 3 
demonstrates that the theoretical results for $\Delta \Phi_A$ lie 
nicely within the bounds provided by the simulation data. However,
quantitative agreement is only reached if the range of monomer interactions
has been accounted for correctly, i.e. $k_2$ may not be neglected.

The total amount of $A$ segregated at the surface is given by the
total excess $Z_A^* = \int dz \rho(z) (\Phi_A(z)-\Phi_{A,b})$. 
As shown in Figure 4, it can be described by a simple Boltzmann
dependence 
\begin{equation}
\label{bb}
Z_A^* \propto \exp(-\chi N \alpha)
\end{equation}
over a wide range of $\chi N$. In this regime, one can describe the 
segregation process as adsorption of single, relatively undistorted $A$ 
chains whose energy is reduced by $\chi N (1-\alpha)$ at the surface,
compared to the bulk. The quantity $(1-\alpha)$ thus gives the quota
of neighbor monomers, which a chain is missing at the surface.
In the present model, one obtains $\alpha = 0.77$. 
Close to the critical point, interactions between different $A$ chains 
at the surface become important and our simple picture 
breaks down - the surface excess diverges. At very high values of $\chi N$, 
beyond  $\chi N {> \atop \sim} 14$, 
the chains are more and more squeezed at the surface, the number
of missing neighbors increases, and the segregation is slightly higher
than eqn (\ref{bb}) would predict.

The surface segregation in the one phase region is shown in Figure 5, as a 
function of bulk volume fraction $\Phi_{A,b}$ at fixed Flory Huggins 
parameter $\chi$. For comparison, segregation profiles for
mixtures of polymers with different chain lengths ($N_B=2 N_A= 48$ or
$N_A=2 N_B=48$ as opposed to $N_B=N_A=32$) were also calculated. In such 
asymmetric mixtures, shorter chains segregate to the surface when there 
are no interactions, at $\chi$ = 0, for entropic reasons (thin line). At 
$\chi {< \atop \sim} \chi_c$, however, the 
segregation is mainly driven by energetic effects: As the bulk volume 
fraction $\Phi_A$ approaches zero, 
polymers $A$ segregate to the surface to the same extent regardless of 
whether they have longer or shorter chain lengths than polymers $B$. The 
effects of segregation due to a difference in chain lengths, and segregation
due to missing neighbor effects, are therefore {\em not} additive. This 
becomes most obvious when looking at the surface excess of component $A$ 
(Figure 5b). On increasing the fraction of chain lengths $N_A/N_B$ while 
keeping $\chi_c = (1/\sqrt{N_A}+1/\sqrt{N_B})^2/2$ approximately constant, 
the critical point moves towards the $A$-poor region of the phase diagram
(cf eqn (\ref{crit})). As a consequence, the excess of $A$-polymers
at the surface is even higher for $N_A > N_B$ than for
$N_B > N_A$. Hence the entropic mechanism which promotes segregation of
the shorter chains is largely suppressed -- the dominant effect of
chain length asymmetry is the shift of the critical point.

\bigskip

\subsection{Structure of the Profiles}

We turn to analyze the segregation profiles in more detail. Far from
the critical point, they can be expected to reflect single
chain behavior. Figure 4 suggests that one may be able to
describe a segregation profile in terms of a weakly adsorbed layer
of the minority component. In order to test this picture, we calculate
for comparison the density profile of a single Gaussian
polymer in  semi-infinite space, which touches the surface at least
once. This can be done analytically, as shown in the appendix. Far from the 
surface, the density profile decays asymptotically like
$\rho(z) \propto 1/z^3 \: \exp(-z^2/(4 R_g^2))$. 
The logarithmic derivative is thus given by 
\begin{equation}
\label{dlra}
R_g \frac{d \log \rho(z)}{dz} = - \frac{3 R_g}{z} - \frac{z}{2 R_g}.
\end{equation}
Figure 6 shows the logarithmic derivative of segregation profiles
$\rho_A(z)$ in the two-phase region at different values of 
$\chi$, compared to the prediction of eqn (\ref{dlra})
(dashed line). Far from the critical point, our simple picture describes
the asymptotic behavior surprisingly well. Close to the critical point, the
logarithmic derivative of $\rho_A(z)$ approaches a constant at large $z$:
The segregation profile decays exponentially with a decay length
$\xi$. This is characteristic for a system close to a critical point. 
The interactions between chains become important, and collective effects 
determine the structure of the profiles. In polymeric systems, the 
characteristic length scale $\xi$ for collective concentration fluctuations 
competes with the gyration radius $R_g$, which describes the correlations 
produced by chain connectivity -- the correlations of monomers belonging to 
the same chain. The result is a crossover between two types of qualitatively 
different asymptotic behavior: Far from the critical point, intrachain
correlations dominate, close by, collective concentration correlations do.
The correlation length $\xi$ diverges at the critical point.

We can again attempt to compare these results with
the simulations of Ref. \cite{rouault}. Simulations where performed at
values $\chi N = 3-4$, where the asymptotic behavior of the profiles
is still governed by the length scale $\xi$. At film
thicknesses of less than $3 R_g$, however, the asymptotic
behavior is never reached. This may explain why the decay length
obtained by Rouault et al, $\xi \approx 0.35 R_g$ at $\chi N = 3.4$, is 
only roughly half the value predicted by the self consistent field theory,
$\xi \approx 0.6 R_g$.

The decay length $\xi$ as a function of $1/\chi N$ is shown in 
Figure 7, and compared to half the width of the bulk interface between
the two coexisting phases, which has been calculated in Ref. \cite{schmid}. 
In a simple system, both lengths are identical, i.e. by the bulk 
correlation length. Here, this is only true very close to the critical
point; further away, $\xi$ gets too large relative to the interfacial
width. This is due to the fact that the
wings of the segregation profiles are governed by the chain end distribution,
as we shall see in the next section. 
The interfacial width, on the other hand, describes the separation of all 
$A$ and $B$ monomers. Chain end distributions tend to be broader than total 
monomer distributions for entropic reasons. In the wings of the interfacial 
profile, we expect to find a decay length which is again 
determined by the chain end distributions and identical to $\xi$.
This is indeed the case (not shown).

Figure 7 also shows data for another length which characterizes
the segregation profiles, the profile width. It is conveniently defined 
through the excess distribution
$Z_A(z) = \rho(z) (\Phi_A(z)-\Phi_{A,b})/\rho_b$, e.g. as its first 
moment or as its standard deviation.
\begin{equation}
W_{\langle z \rangle} = \frac{\int dz\: z Z_A(z)}{Z_A^*} \qquad 
W_{\langle \delta z^2 \rangle} = \sqrt{
\frac{\int dz\: z^2 Z_A(z)}{Z_A^*} - (\frac{\int dz\: z Z_A(z)}{Z_A^*})^2}
\end{equation}
At the critical point, the width diverges with the correlation length $\xi$. 
Further away, it reaches a plateau at $W \approx R_g/\sqrt{3}$, which
is about the extention of an undistorted chain in one dimension.
At large values of $\chi N$, however, the width decreases further and 
approaches zero. The strong repulsion between chains $A$ and $B$ cause
the segregated chains to get increasingly distorted and squeezed towards 
the surface. The turning point of the curve is at $1/\chi N \approx 0.07$
or $\chi N \approx 14$, consistent with what was already deduced from the
$\chi N$--dependence of the surface excess $Z_A^*$ in Figure 4.

\bigskip

\subsection{Chain end distributions and Chain orientations}

Figure 8 compares the distribution of chain ends with density profiles close
to the surface at bulk two phase coexistence. For both the minority and
the majority component, the fraction of chain ends relative to the total
density is augmented right at the surface, and reduced in a region underneath. 
Such an effect is already known from pure systems\cite{kumar,wang1}. In the 
minority component, the depletion zone is followed by
another enrichment zone in the wings of the decaying segregation profile --
polymers of the minority component rather stick out chain ends into the 
majority phase than whole loops. 
Thus the segregated layer of polymers $A$ is skirted by regions of 
enhancement of chain ends $A$. The total chain end distribution 
profile, however, only has one enrichment zone right at the surface. 
One can probably expect a second zone at higher segregation,
beyond the wetting transition, since chain ends are known 
to enrich at the $A$/$B$ interface \cite{marcus2,schmid}.
In our simple picture of an adsorbed polymer layer, the chain end distribution 
decays asymptotically like $\rho(z) \propto 1/z \: \exp(-z^2/(4 R_g^2))$
(see appendix), i.e. the logarithmic derivative follows
\begin{equation}
R_g \frac{d \log \rho_{end}(z)}{dz} = - \frac{R_g}{z} - \frac{z}{2 R_g}.
\end{equation}
This is compared to the logarithmic derivative of chain end distributions
of the segregating component at different values of $\chi$ in Figure 9. 
The agreement for high values of $\chi N$ is not as good as for the 
total density profiles, but still remarkable. Close to the critical point,
the profiles show asymptotically again exponential decay with
the decay length $\xi$.

The orientation of whole chains can be studied by solving the diffusion
equation (\ref{qq}) in the previously determined self consistent field
$W_i$ with initial condition $Q_i(\vec{r},0) = \delta(\vec{r}-\vec{r}_0)$. 
The statistical weight of a polymer with end-to-end vector 
$\vec{R}_e = \vec{r}_1-\vec{r}_0$ and center at 
$\vec{R}_c = (\vec{r}_1 + \vec{r}_0)/2$ is then given by $Q_i(\vec{r}_1,0)$. 
Since the self consistent field $W_i$ only varies in the direction 
perpendicular to the interface, $z$, the components of the end-to-end vector
parallel to the interface have the same distribution at the surface than in 
the bulk. The presence of a surface only affects the $z$-component $R_{z,e}$.
Chains get therefore oriented as a result of stretching or squashing in the 
$z$ direction, the theory does not allow for the possibility of chain
orientation without compression, as has been observed in computer
simulations\cite{kumar}.
Profiles of $\langle R_{e,z}{}^2\rangle(z)$, where the averages have been
performed for polymers with center at $z$, are shown in Figure 10.
The bulk value of this quantity is simply $2 R_g{}^2$
with the gyration radius $R_g$. Near the surface, it is greatly
reduced -- it drops down to zero at $z=0$ and remains smaller than the
bulk value over a distance of more than one gyration radius from the surface. 
Hence polymers tend to align parallel to the surface. This region is followed
by a zone of perpendicular alignment at the distance $2 R_g$ from the surface 
(Figure 10). Both the parallel alignment\cite{yethiraj} and the subsequent 
``overshoot'' of perpendicular alignment\cite{wang1}
have been seen in Monte Carlo simulations of on one component polymer melts.
The latter effect is weak for chains of the majority component, but 
can be quite marked for the minority component, deep in the two-phase region. 
The extent of parallel alignment, on the other hand, does not 
depend on whether a chain belongs to the minority or the majority component, 
or on the strength of the interaction $\chi$. 
Thus we find that the total structure of the blend does not change with
increasing $\chi$, although the segregated minority component
chains get more and more distorted, as indicated by Fig. 7. They just
accumulate closer to the surface, where the distortion is higher.

\bigskip

\section{Summary}

\bigskip

A self consistent field theory for compressible binary
polymer mixtures has been developed, which allows for a consistent
treatment of surfaces, including effects of missing neighbors.
It has been shown to be remarkably successful in reproducing Monte Carlo 
results for the segregation at a neutral surface in the bulk 
coexistence region, without resorting to any fitting parameter.
On this base, several aspects of the surface structure
could be discussed, like the role of chain end distribution and chain 
orientations, the effect of chain length asymmetry, 
and the interplay of the different length scales which govern the system: 
the correlation length of collective concentration fluctuations and 
the single chain correlation length, i.e. the gyration radius. 
Hence we have seen that the formalism developed in this work is suited
for the investigation of surfaces of polymer mixtures. So far it has
only been applied to systems with perfectly symmetric interactions
(where  $V_3(\vec{r}) \propto (U_{AA}(\vec{r}) - U_{BB}(\vec{r}))$ 
(eqn (\ref{fi2})) vanishes), a rather unlikely situation in real systems.
Usually the asymmetry of the interactions introduces 
additional surface potentials $h(z)$ (cf. eqn (\ref{fields}), i.e.
preferential attraction of one component to the surface, 
After taking those into account, the theory should be similarly successful
in describing experimental results.

I wish to thank K. Binder for suggesting the problem and for useful 
comments on the manuscript, and L. Klushin and O. Borisov for helpful
discussions.

\bigskip

\section*{Appendix}

\bigskip

We consider a Gaussian Polymer of length $N$ in the half space with
end monomers located at distances $z_0$ and $z_1$ from the surface,
which touches the surface with the $K$th monomer (Figure 11).
The Greens function for propagation in the half space can be obtained
with the method of mirror images 
\begin{equation}
G(M,z;M',z') \propto
\exp(-\frac{(z-z')^2}{\frac{2}{3}(M'-M)b^2}) -
\exp(-\frac{(z+z')^2}{\frac{2}{3}(M'-M)b^2}),
\end{equation}
where $b$ is the Kuhn length, and $z$, $z'$ are the locations of monomer
$M$ and $M'$ \cite{wiegel}.
Hence the distribution profile of monomer $L$ is given by
\begin{equation}
D_L(z|0,z_0;K,\epsilon \to 0^+;N,z_1) 
\propto \hspace*{6cm}
\end{equation}
\begin{displaymath}
\lim_{\epsilon \to 0^+}
\left\{ 
\begin{array}{lcr}
G(0,z_0;L,z) G(L,z;K,\epsilon) G(K,\epsilon;N,z_1) &:& K>L \\
G(0,z_0;K,\epsilon) G(K,\epsilon;L,z) G(L,z;N,z_1) &:& L>K 
\end{array}. \right.
\end{displaymath}
In order to get the total density of the $L$th monomer, one has to integrate
over the free parameters $z_0$, $z_1$ and $K$. The result is
\begin{equation}
\rho_L(z) \propto g_l(\alpha z) \sqrt{l} \sqrt{1-l}
\end{equation}
with
\begin{displaymath}
g_l(x) = - x \bigg[
\Phi(\frac{x}{\sqrt{l}})\mbox{Ei}(-\frac{x^2}{1-l})\frac{1}{\sqrt{1-l}} +
\Phi(\frac{x}{\sqrt{1-l}})\mbox{Ei}(-\frac{x^2}{l})\frac{1}{\sqrt{l}}
\bigg],
\end{displaymath}
where $l=L/N$ and $\alpha = 1/(2 R_g)$ with the gyration radius
$R_g^2 = N b^2/6$. The error function $\Phi$ is defined as
$\Phi(x) = \frac{2}{\sqrt{\pi}}\int_0^x \exp(-t^2) \: dt$, and the
exponential integral Ei as
$-\mbox{Ei}(-x) = \int_x^{\infty} \exp(-t)/t \: dt$. Note that the 
normalization factor ${\cal N}(\sqrt{l}) = \int_0^{\infty} dx g_l(x)$
is finite and nonzero for all values of $l$ between zero and one. 
The total density profile is given by
\begin{equation}
\rho(z) = \int_0^1 dl \: g_l(\alpha z)/{\cal N}(\sqrt{l}).
\end{equation}
The asymptotic behavior of the profiles can be obtained from the
asymptotic behavior of $g_l(x)$:
\begin{equation}
g_l(x) \stackrel{x \to \infty}{\longrightarrow}
\frac{1}{x} [\: \exp (-\frac{x^2}{1-l}) \sqrt{1-l} 
+  \exp (-\frac{x^2}{l}) \sqrt{l} \:].
\end{equation}
The result for the chain end distribution follows immediately.
\begin{equation}
\rho_0(z) = \rho_1(z) \propto
\frac{1}{z} \exp(-(\alpha z)^2)
\end{equation}
The asymptotic behavior of $\rho(z)$ is slightly more difficult to derive. 
Wihout calculating ${\cal N}^{-1}(y)$ explicitly, we expand it in powers
of $y$: ${\cal N}^{-1}(y) = \sum_k c_k y^k$. Using asymptotic series
representations of the error function and the exponential integral,
one can show
\begin{equation}
\int_0^1 dl \: l^{k/2} e^{-x^2/l} 
\stackrel{x \to \infty}{\longrightarrow}
e^{-x^2} \frac{1}{x^2} \sum_{j=0}^{N} (-1)^j
\frac{\Gamma(j + k/2)}{\Gamma(k/2)} (\frac{1}{x^2})^j 
+ {\cal R}_N,
\end{equation}
with the Gamma function $\Gamma$ and 
${\cal R}_N < x^{-2(N+1)} \Gamma(N)$. With that one obtains
\begin{eqnarray}
\rho(z) &\approx& \frac{2}{\alpha z} \sum_{k=0}^{\infty} c_k
\int_0^1 dl \: k^{(k+1)/2} e^{-(\alpha z)^2/l} \nonumber\\
&\approx& 
\frac{2}{(\alpha z)^3} e^{-(\alpha z)^2}
\sum_{k=0}^{\infty} c_k \sum_{j=0}^N (-1)^j
\frac{\Gamma(j+(k+1)/2)}{\Gamma((k+1)/2)} (\frac{1}{x^2})^j \nonumber\\
& \longrightarrow &
2 c_0 e^{-(\alpha z)^2} \frac{1}{(\alpha z)^3}
\end{eqnarray}

\newpage
\section*{Figure Captions}

\begin{description}

\item[Figure 1:] 
Mean field phase diagram of the symmetric polymer blend with $N_A = N_B=N$
in the plane of variables $1/(\chi N)$ and the volume fraction $\Phi_A$ of
$A$ monomers. Inside the binodal curve (solid line), the mixture
phase separates into two macroscopic phases (1) (left) and (2) (right).
Also shown is the surface volume fraction of $A$ monomers at a free
surface $\Phi_{1,A}$ in each of the two coexisting phases. 
Near the critical point, the concentration differences 
$(\Phi_A^{(1)}-\Phi_{A}^{(2)})$ and $(\Phi_{A,1}^{(1)}-\Phi_{A,1}^{(2)})$
vanish with different critical exponents: 
$(\Phi_A^{(1)}-\Phi_{A}^{(2)})\propto (\chi_c-\chi)^{\beta}$ with $\beta=1/2$
and $(\Phi_{A,1}^{(1)}-\Phi_{A,1}^{(2)})\propto (\chi_c-\chi)^{\beta_1}$
with $\beta_1=1$ in mean field theory. Thus the broken curves have a
cusp at the critical point.

\item[Figure 2:]
Volume fraction profile of the minority component monomers at 
$\chi/\chi_c = 1.696$ in the symmetric polymer mixture. 
Dashed line shows volume fraction profile for $k_2=0$, where
monomers are assumed to have pure contact interactions with
each other. Inset shows the total density profile, which
was identical for all considered values of $\chi$.

\item[Figure 3:]
Difference $\Delta \Phi_A$ between surface and bulk volume fraction of the
minority component monomers in the symmetric polymer mixture, at two
phase coexistence,
vs. $1/(\chi N)$. Data points show results of the simulations of
Rouault et al (\cite{rouault}): The difference between surface and bulk
volume fraction (circles), and the difference between the surface
volume fraction and the volume fraction at the center of simulated slab
(squares). The arrow indicates the value of $1/(\chi N)$ corresponding
to the profiles in Figs 2 and 8.

\item[Figure 4:]
Surface excess $Z_A^*$ of the minority component at two phase coexistence in 
the symmetric polymer mixture, in units of $\rho_b R_g$, vs $\chi N$.
dashed line shows the function $0.3 \exp(-0.77 \chi N)$.

\item[Figure 5:]
(a) Difference $\Delta \Phi_A$ between surface and bulk volume fraction of 
$A$ monomers and (b) surface excess $Z_A^*$ in units of $\rho_b R_g$,
in the one phase region, as a function of the bulk volume fraction
Considered are symmetric mixtures with $N_A=N_B$
(thick solid line) and asymmetric mixtures, where $A$ polymer chains are either
longer (long dashed line) or shorter (short dashed line) than
$B$ polymer chains. Thin line indicates results for asymmetric mixtures
with the interactions turned out or $\chi=0$.

\item[Figure 6:]
Logarithmic derivative of density profiles for the minority component
in the symmetrical mixture, in units of $R_g^{-1}$, at two phase coexistence.
Values of $\chi/\chi_c$ are, from top to bottom:\\
$\chi/\chi_c = \{1.102, 1.271, 1.695, 2.542, 3.390, 4.237, 
5.085, 5.932, 6.780, 7.627, 8.475 \}$. Dashed line shows 
the function $-z/(2 R_g) - (3 R_g)/z$ for comparison (see text for
explanation).

\item[Figure 7:]
Characteristic lengths of the segregation profiles in units of the gyration
radius, vs $1/(\chi N)$, in the symmetrical mixture at two phase coexistence. 
Solid line and long dashed line show the width of the profiles, defined in 
two different ways. Short dashed line gives asymptotic decay length $\xi$, 
dotted line gives half the interfacial width between the two coexisting 
phases in the bulk. Dashed dotted line indicates the extension of one 
undistorted chain in the bulk. Triangles mark the values of
$1/(\chi N)$, corresponding to the profiles shown in Figs 6 and 9.

\item[Figure 8:]
Distribution of chain end monomers of the minority component $\rho_{A,end}$ 
in units of $(2 \rho_b/N)$ in the symmetric polymer mixture at two phase
coexistence, $\chi/\chi_c = 1.696$. Dashed line indicates distribution 
of all monomers for comparison. Inset shows the total chain end distribution 
(solid line), compared to the total density profile (dashed line).

\item[Figure 9:]
Logarithmic derivative of chain end distributions for the minority component
in the symmetrical mixture, in units of $R_g^{-1}$, at two phase coexistence.
Values of $\chi/\chi_c$ are as in Figure 6, from top to bottom: \\
$\chi/\chi_c = \{1.102, 1.271, 1.695, 2.542, 3.390, 4.237, 
5.085, 5.932, 6.780, 7.627, 8.475 \}$. Dashed line shows 
the function $-z/(2 R_g) - R_g/z$ 

\item[Figure 10:]
Profiles of the average squared $z$-component of the end-to-end vector
$\langle R_{e,z}^2 \rangle$ for chains centered at $z/R_g$, vs
$z/R_g$. Considered are chains of the minority component (dashed lines) and 
the majority component (solid line) in a symmetrical mixture at two phase 
coexistence, different values of $\chi$. The results for chains of the majority
component do not depend on $\chi$.

\item[Figure 11:]
Cartoon of a single weakly adsorbed chain in semi-infinite space

\end{description}


\begin{thebibliography}{99}
\bibitem{comment}
Note that this procedure is not unambiguuous, since $\alpha$ and $\beta$ 
can be interchanged. Hence there is some arbitrariness in the exact 
expression for the total free energy, eqn. (\ref{free}). This is however
not a problem, since the expressions for the mean fields, eqn. (\ref{fields}),
remain unequivocal.
\end{thebibliography}
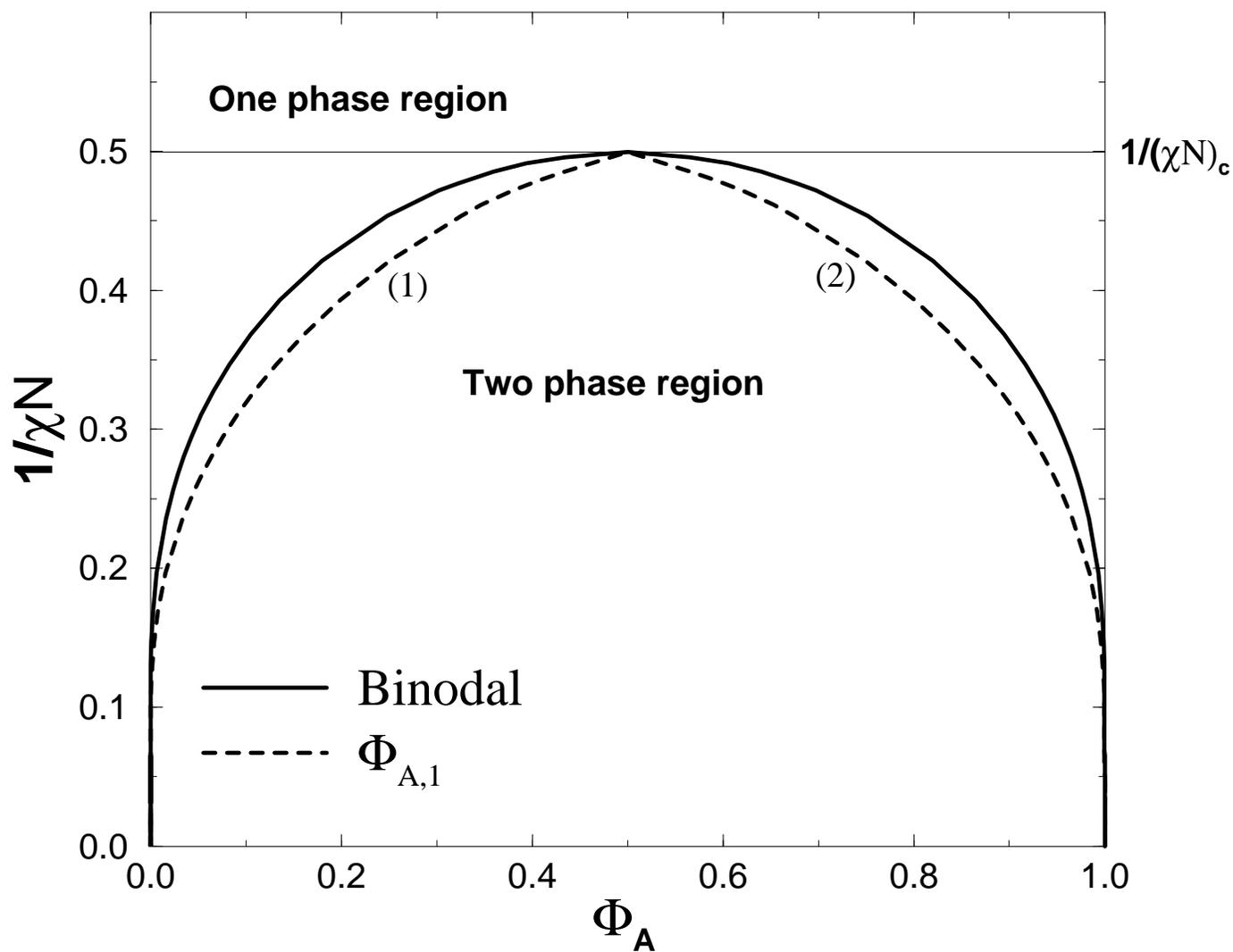
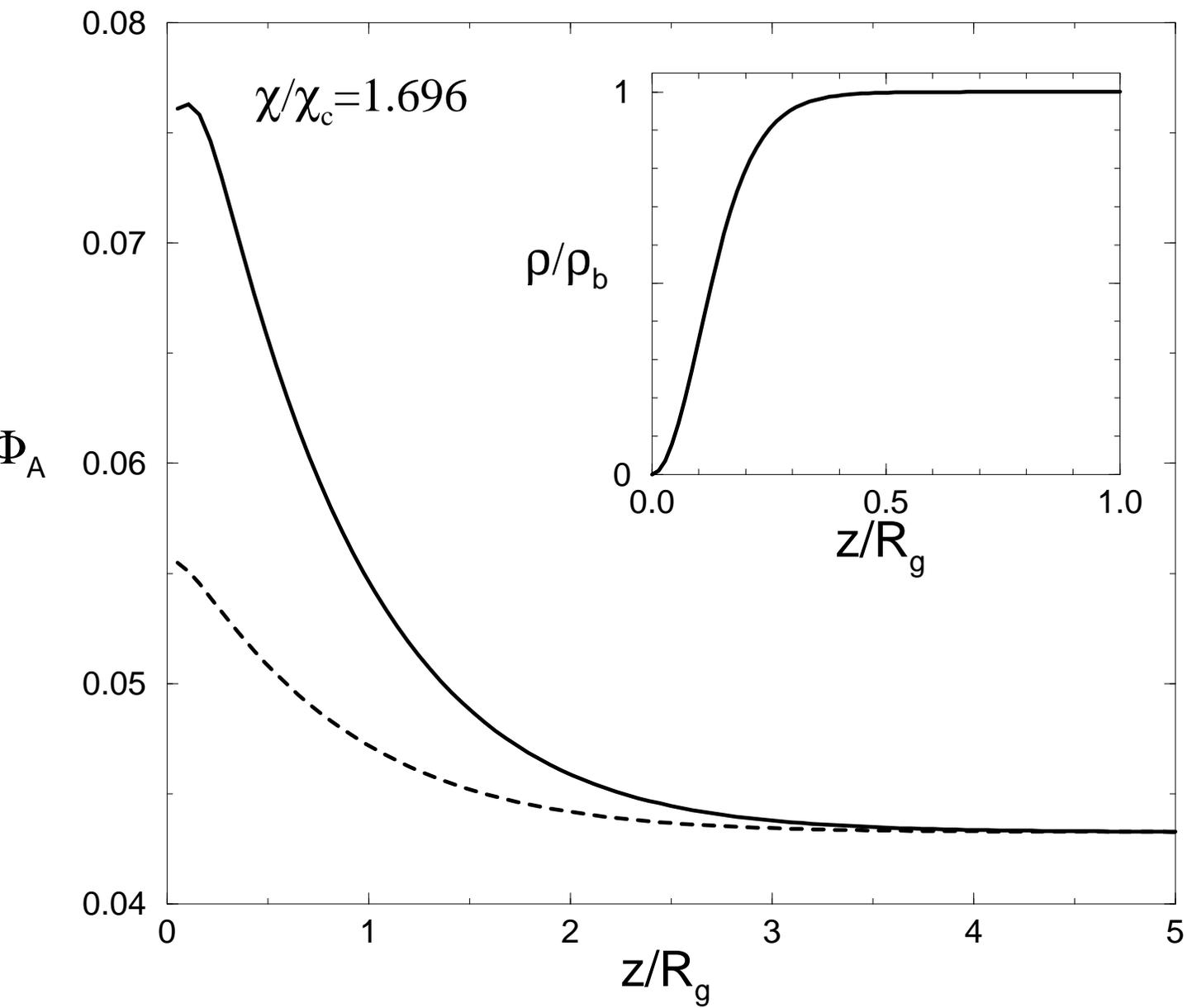
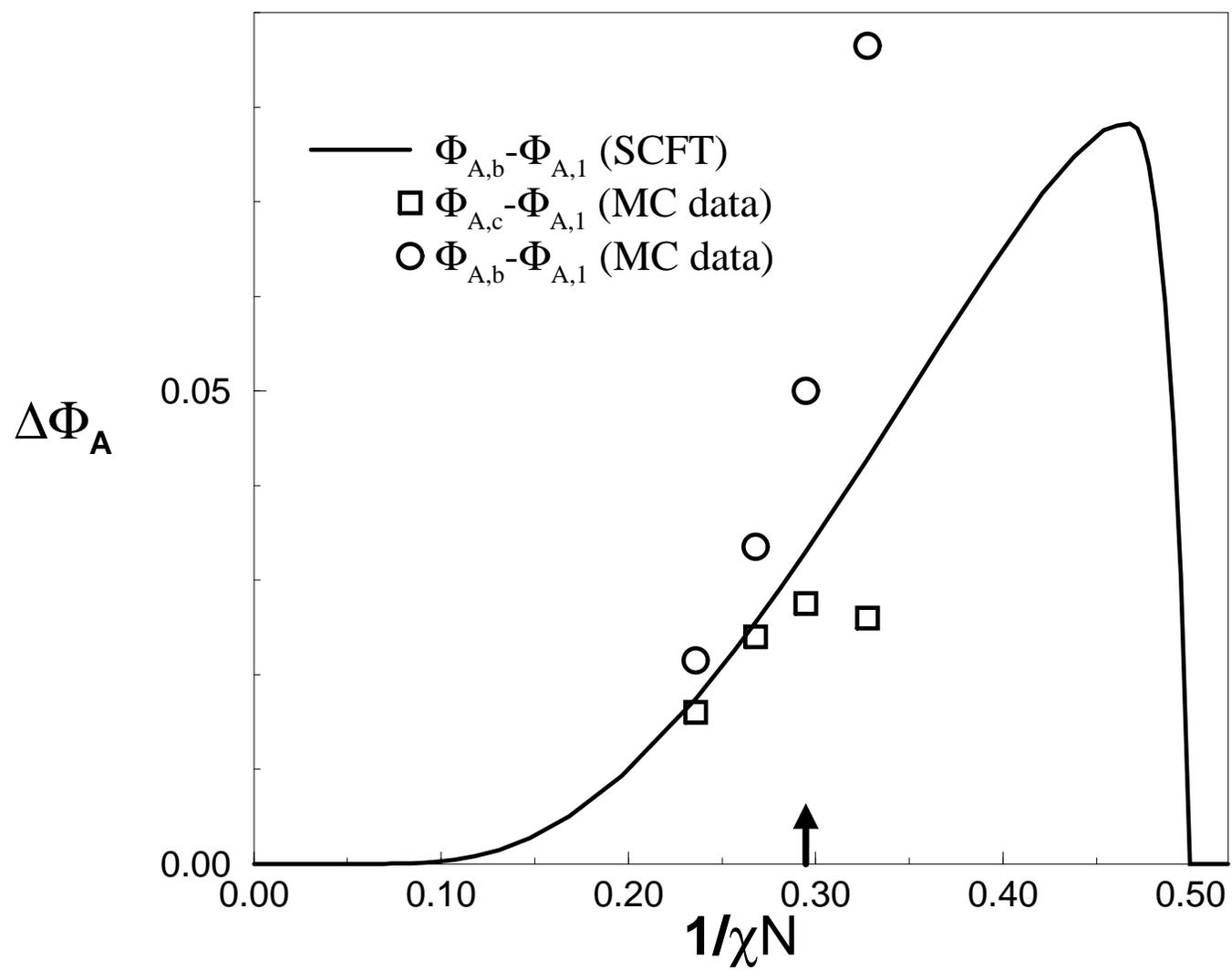
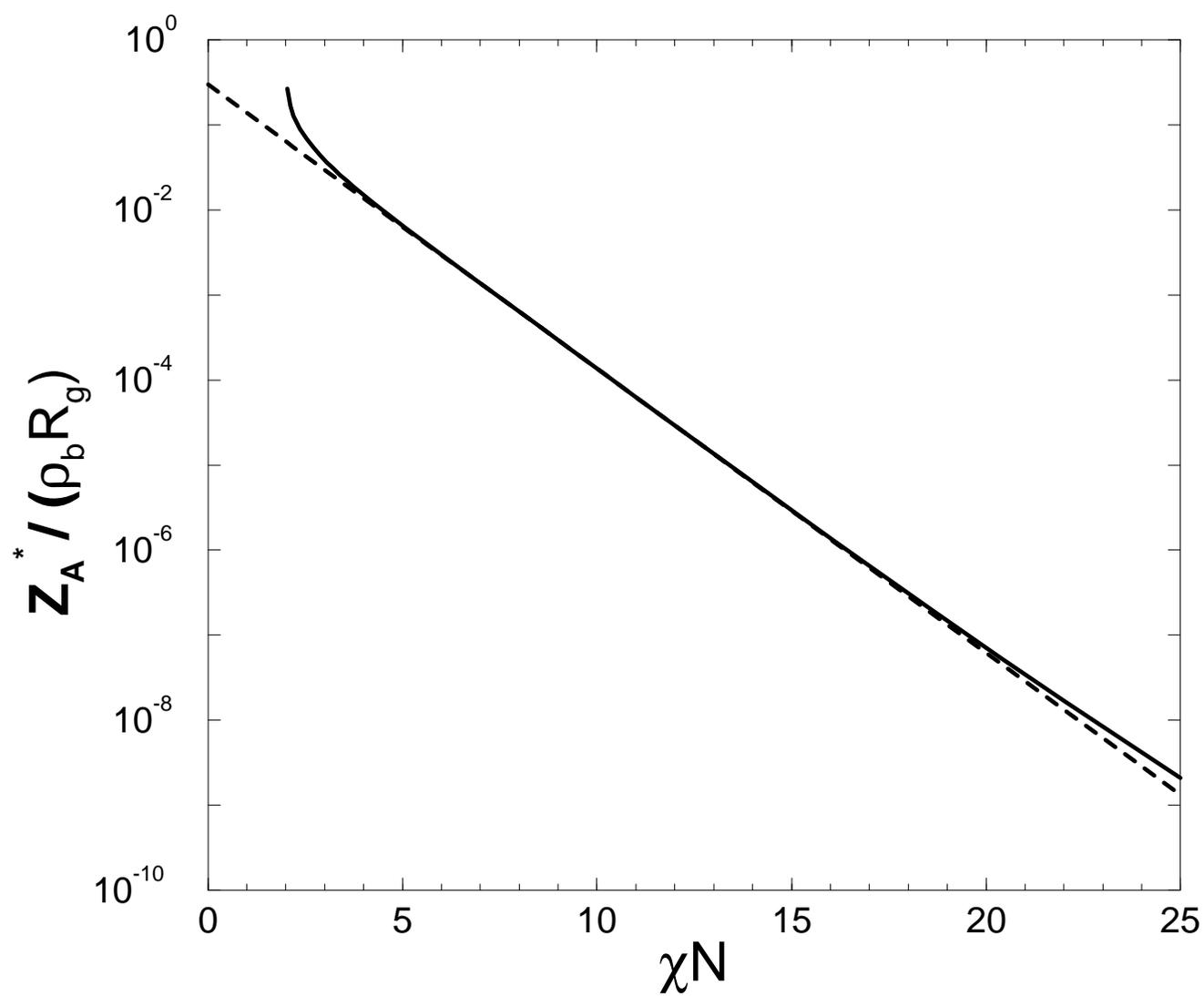
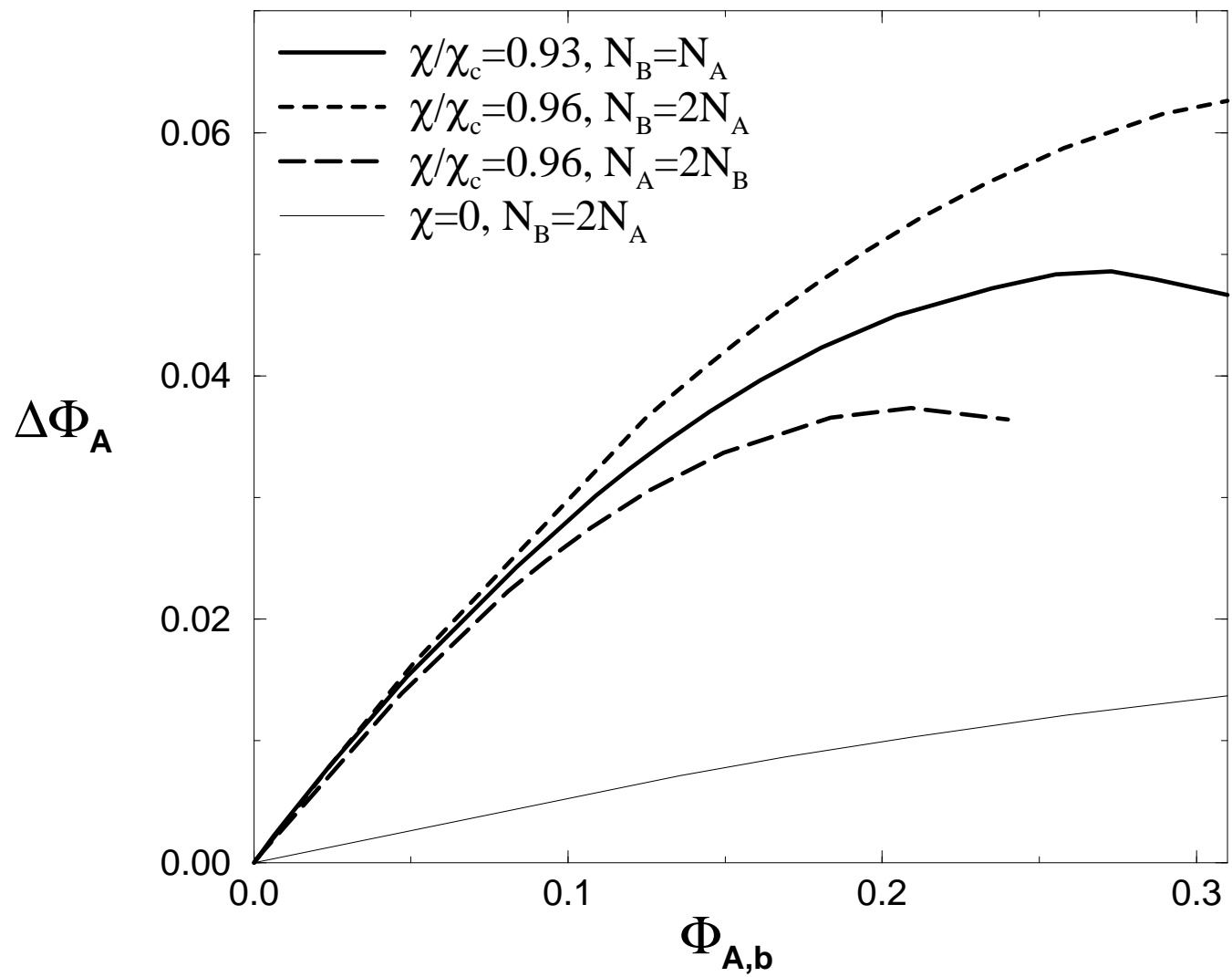
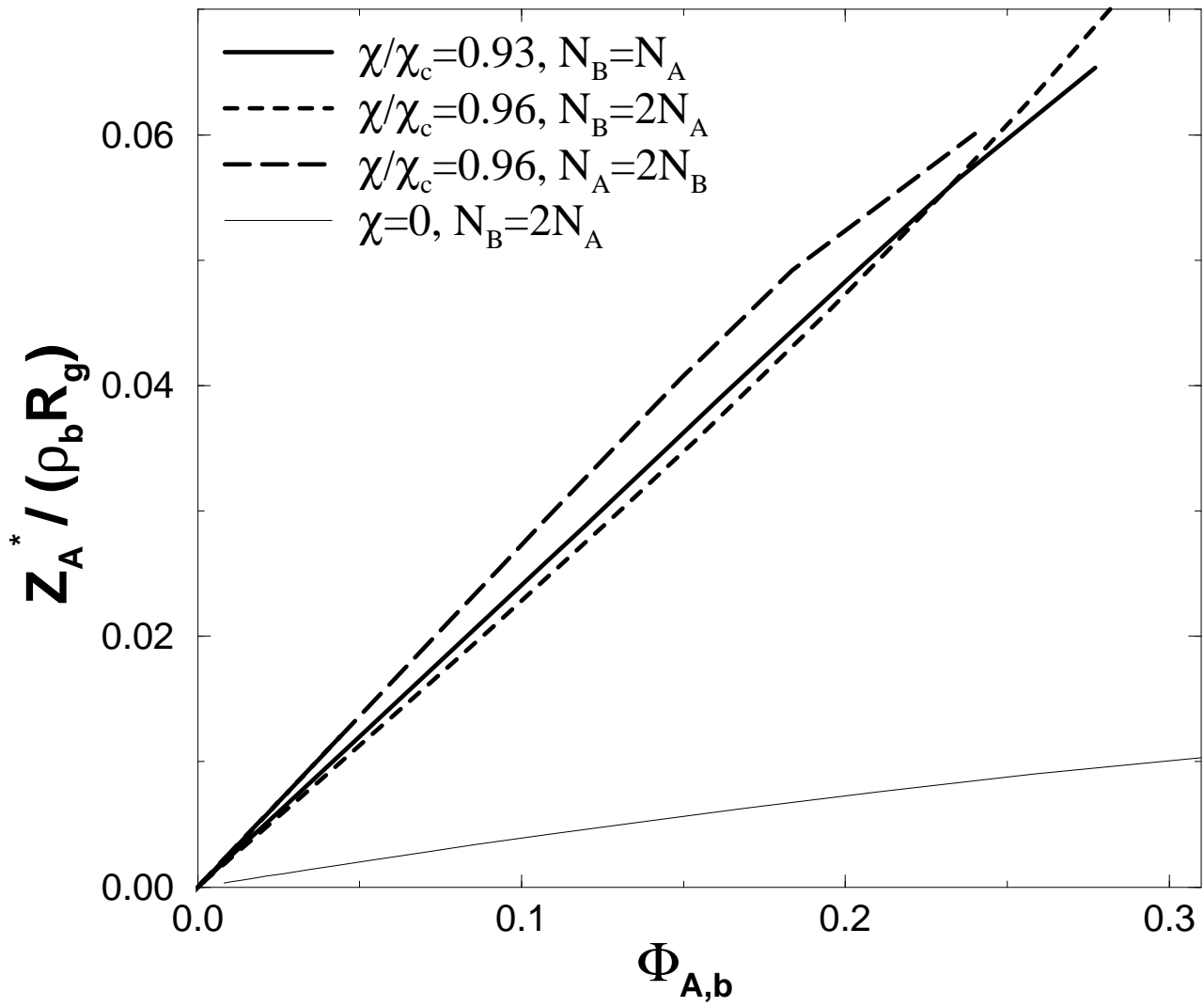
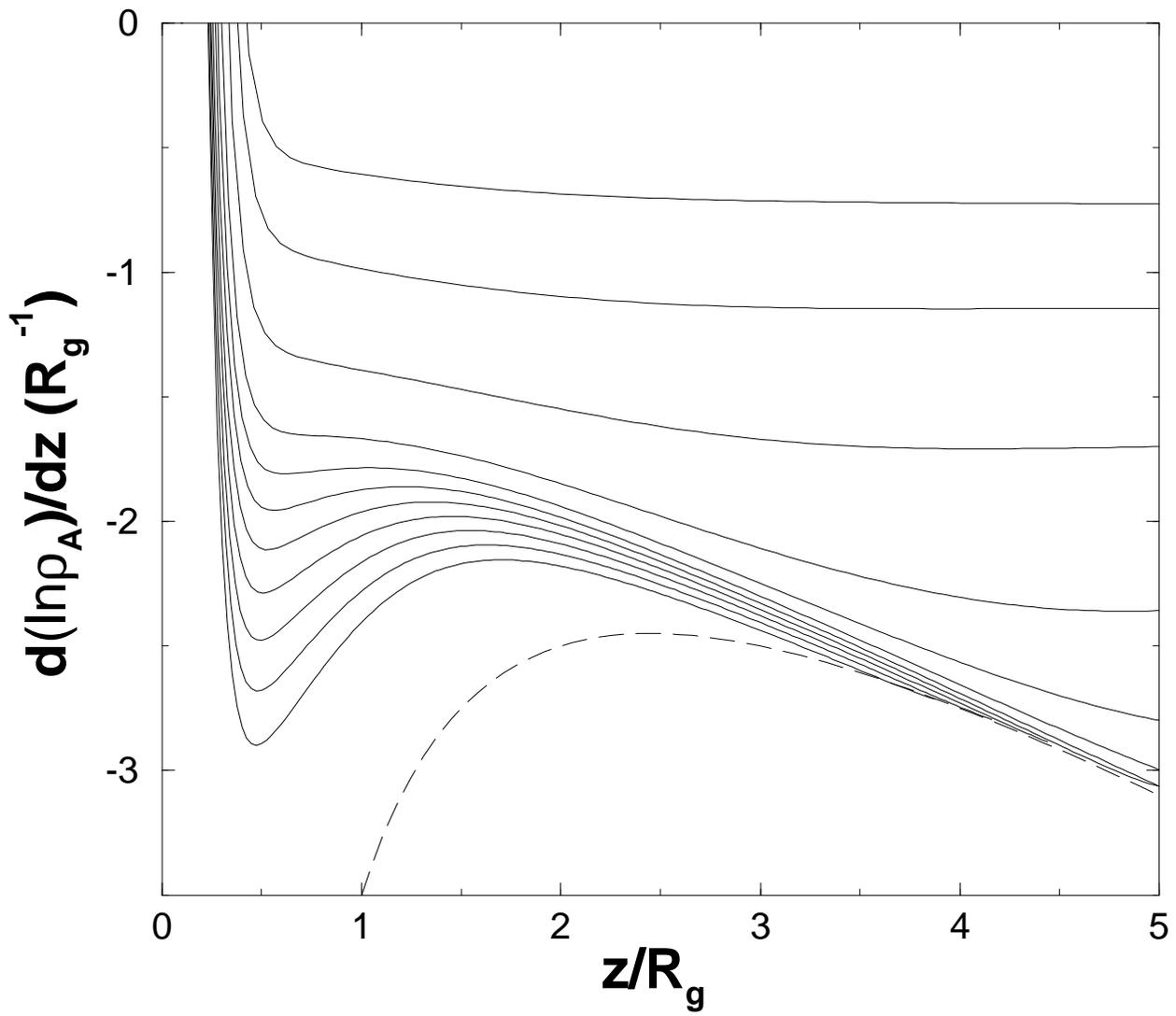
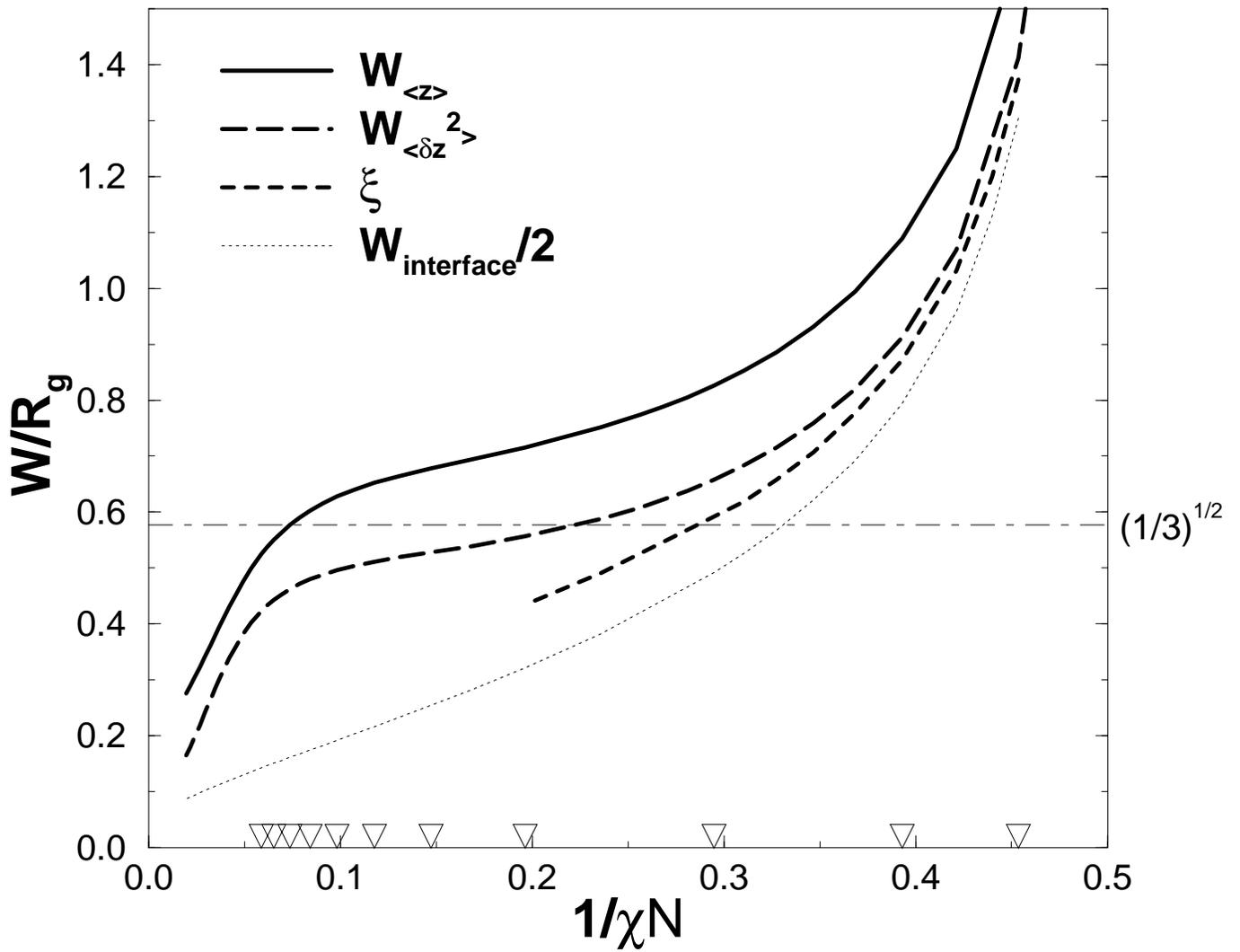
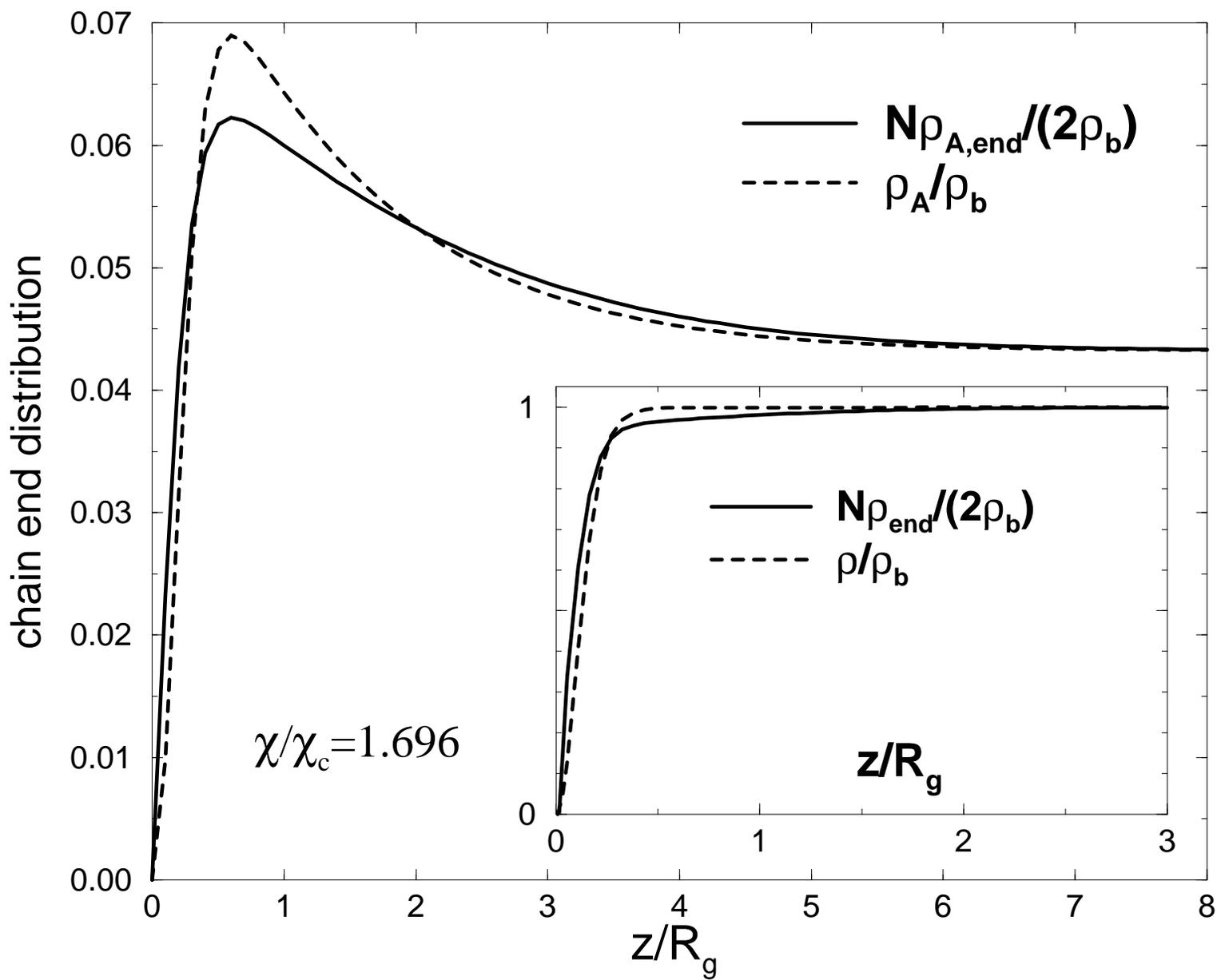
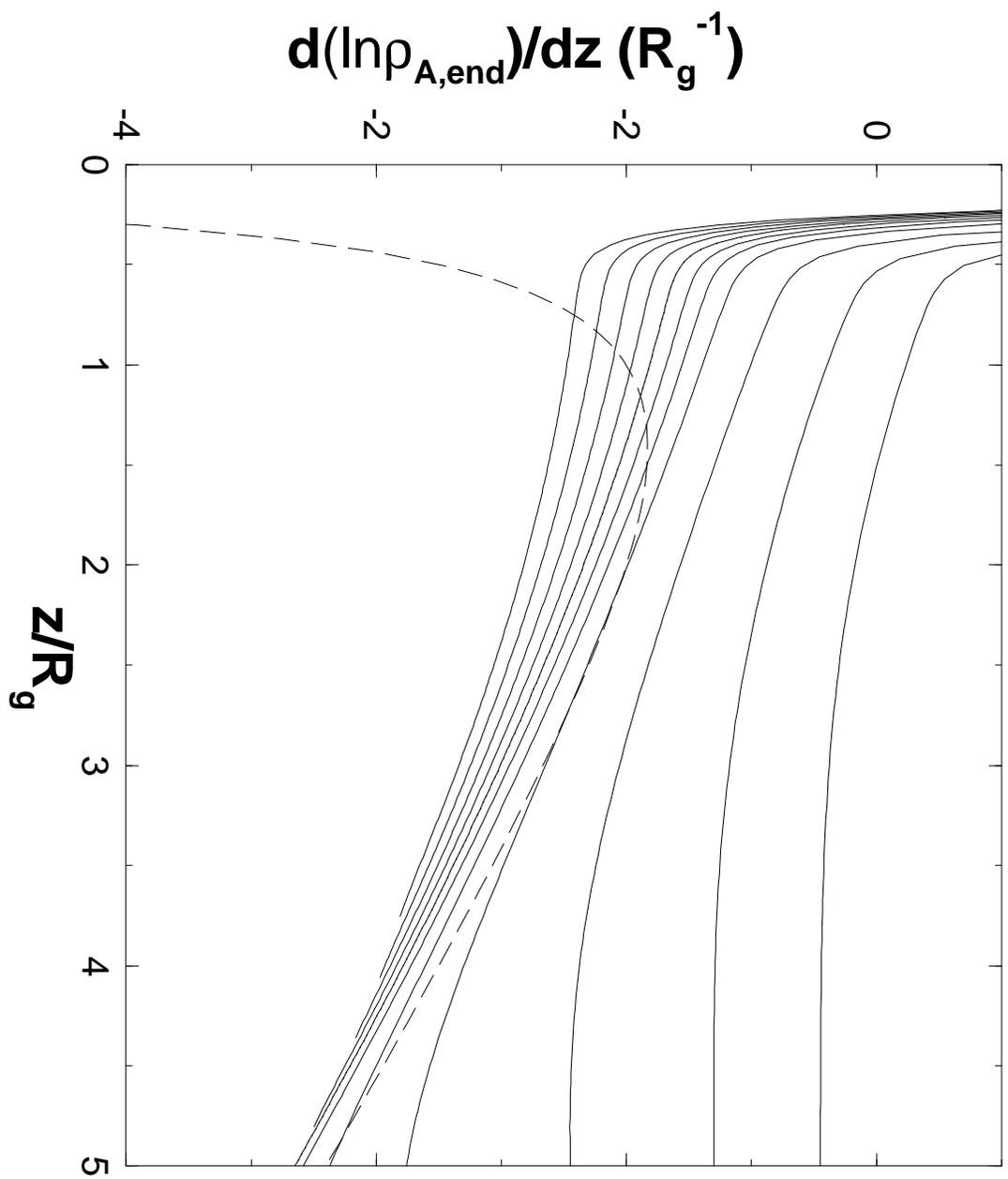
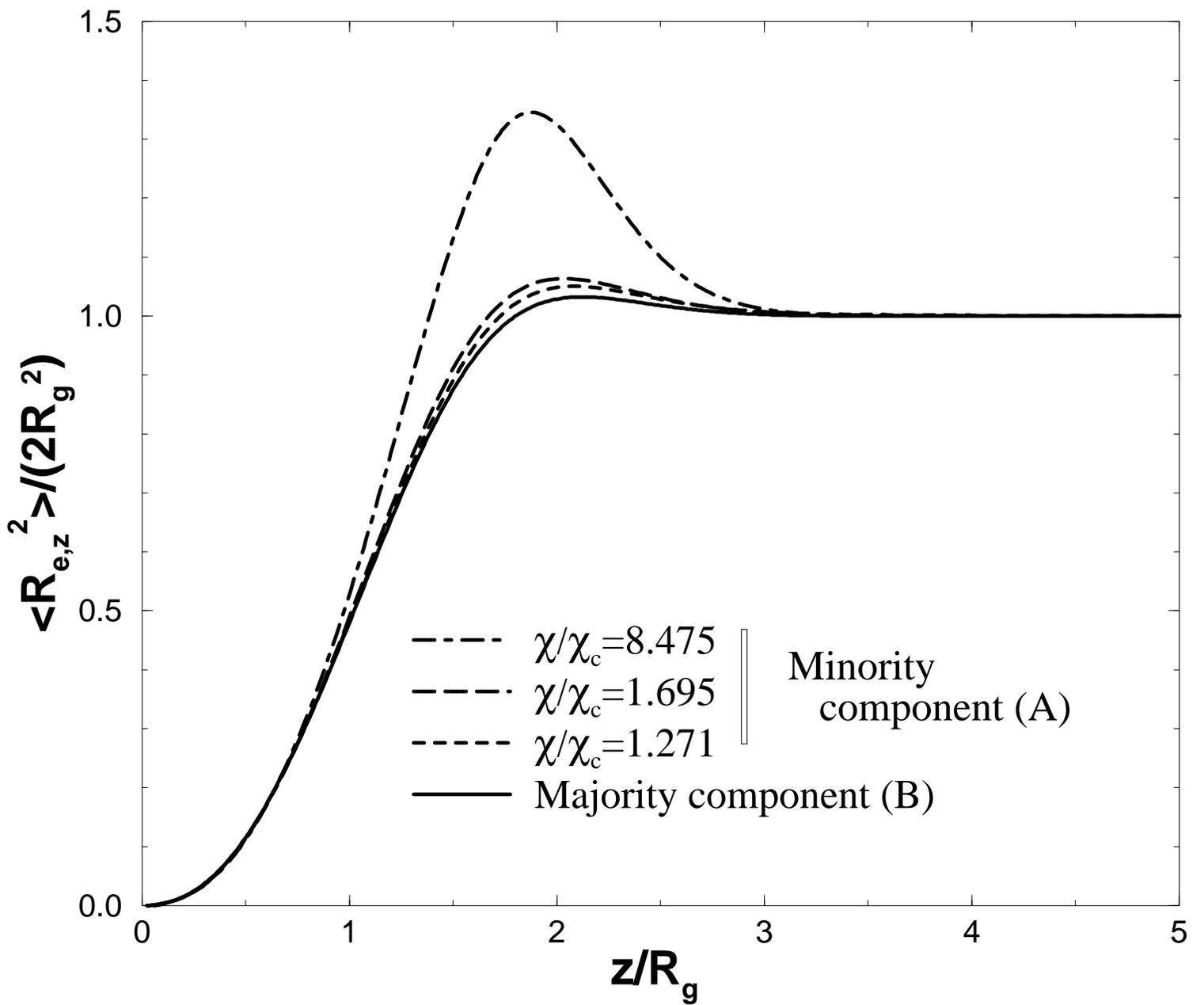
\end{document}